# Quantitative analysis of intermediately and infinitely thick samples with thin sample approach without sample preparation using confocal X-ray fluorescence*

SUN Xue-Peng (孙学鹏) [1,2,3], LIU Zhi-Guo (刘志国) [1,2,3], SUN Tian-Xi (孙天希) [1,2,3,†]

Peng Song (彭松) [5], Sun Wei-Yuan (孙蔚渊) [1,2,3], Li Fang-Zuo (李坊佐) [1,2,3]

Jiang Bo-Wen (姜博文) [1,2,3], Ma Yong-Zhong (马永忠) [4], Ding Xun-Liang (丁训良) [1,2,3]

[1] The Key Laboratory of Beam Technology and Materials Modification of the Ministry of Education, Beijing Normal University, Beijing 100875, China

[2] College of Nuclear Science and Technology, Beijing Normal University, Beijing 100875, China

[3] Beijing Radiation Center, Beijing 100875, China

[4] Center for Disease Control and Prevention of Beijing, Beijing 100013, China

[5] General Research Institute for Nonferrous Metals, Beijing 100088, China

**ABSTRACT:** In order to validate that the confocal X-ray fluorescence had potential applications in analyzing the intermediately and infinitely thick samples with thin sample approach without sample preparations, as an example, the confocal X-ray fluorescence based on polycapillary X-ray optics was used to analyze multi elements solutions. A model of "equivalent thin sample" formed by the confocal micro-volume overlapped by the foci of the polycapillary X-ray optics was developed. The sensitivity factors of elements applied to the developed method were measured. The nondestructive quantitative analysis of a group of multi-elements solutions with different concentration was performed

*This work was supported by the National Natural Science Foundation of China (11375027) and the Fundamental Research Funds for the Central Universities (2014kJJCA03)
† E-mail: stx@bnu.edu.cn

## 1 Introduction

X-ray fluorescence (XRF) is an important nondestructive method of determining the elemental composition for materials, biological and environmental specimens, food, medicine, and so on [1-5]. The approach to quantification in XRF analysis is different for thin, intermediately and infinitely thick samples. A major feature of the thin sample method is that the intensity of characteristic X-rays depends linearly on the concentration of the corresponding element. In other word, the so-called matrix effects can safely be neglected [6]. For both intermediately and infinitely thick sample methods, the matrix effects should not be neglected. Although there are many methods which can be used to correct the matrix effects, such uncertain matrix effects cannot be corrected fully [7, 8]. Therefore, the errors of both intermediately and infinitely thick sample methods are generally lager than that of the thin sample method. Although the thin sample method has smaller errors than other methods, the process of its sample preparation is complicated. In order to avoid this complicated sample preparation, in this paper, a confocal XRF was used in the quantitative analysis of intermediately and infinitely thick samples using thin sample approach without sample preparation.

The confocal XRF can base on such X-ray optics as Kirkpatrick-Baez mirrors, compound refractive lenses, capillary X-ray optics, and so on [9-12]. In recent years, the confocal technology based on polycapellary X-ray optics is popular. The polycapillary X-ray optics, also known as the Kumakhov lens, works on the total external reflection and can be used to collect the X-rays in wide energy band from a relatively large source angle [13]. The polycapillary X-ray optics can be divided into two main types of polycapillary focusing X-ray lenses (PFXRL) and polycapillary parallel X-ray lenses (PPXRL). The PFXRL can be used to focus the divergent or quasi-parallel X-ray beam to an output focal spot (OFS) with a diameter of about 30 μm and a magnitude of the gain in power density of $10^3$. The PPXRL can be used to collect the X-rays from a finite region which is called input focal spot (IFS) of the PPXRL. When a PFXRL in the excitation channel and a PPXRL coupled with a detector in the detection channel are in a confocal configuration, the overlap of the OFS of the PFXRL and the IFS of the PPXRL, a probing volume, can be formed and only the X-rays emitting from this specific volume can be detected. By successively moving sample located at the confocal position, three-dimensionally resolved information of the sample can be obtained. Considering this feature of the confocal technology, it is widely used in confocal XRF, confocal X-ray absorption fine structure, confocal X-ray diffraction, confocal small-angle X-ray scattering and full-field transmission X-ray imaging technology [14-20]. For the quantitative analysis of sample using the confocal XRF based on the polycapellary X-ray optics, a few methods have so far been developed. The fundamental parameter approach assuming a spherical probing volume has been proposed for the analysis of paint layers [21]. A more detailed model of confocal volume has been proposed and a general equation for the depth-dependent intensity of XRF radiation in confocal geometry as well as a calibration procedure have been derived by Malzer and Kanngieβer [22]. A quantification approach based on a Monte Carlo simulation code has been presented by Czyzycki [23].

In order to validate that the confocal XRF had potential applications in analyzing the intermediately and infinitely thick samples with thin sample approach without sample preparations, the confocal XRF based on the polycapillary X-ray optics was used to determine the concentration of the multiple ions in solutions with thin sample approach.

**2 Experimental setup and methods**

**2.1 Confocal XRF setup**

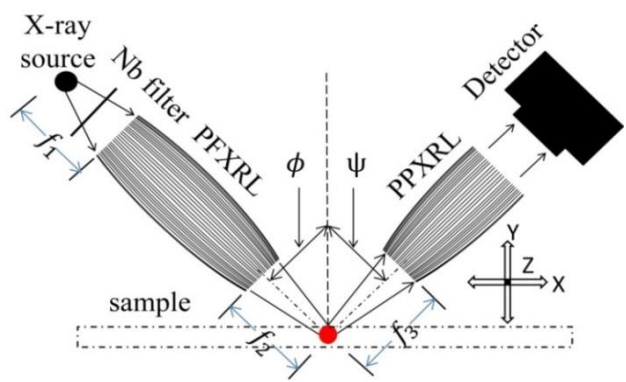

Fig. 1. Scheme of the confocal XRF setup.

The sketch of the confocal XRF from top view is shown in Fig. 1. The X-ray source was a Mo rotating anode X-ray generator with a source spot size of about 300×300 μm². It was placed at the IFS of the PFXRL with a distance of $f_1$ away from the entrance of the PFXRL. An Nb filter was placed between the source and the PFXRL to acquire quasi-monochromatic incident X-rays. The PPXRL was placed confocally with the PFXRL, and $f_2$ and $f_3$ were the output focal distance of the PFXRL and input focal distance of the PPXRL, respectively. The PFXRL, sample and PPXRL coupled with a detector were adjusted by five-dimensional (5D) stage, respectively. [24] The energy resolution of the detector used in the experiment was about 140 eV at 5.9 keV.

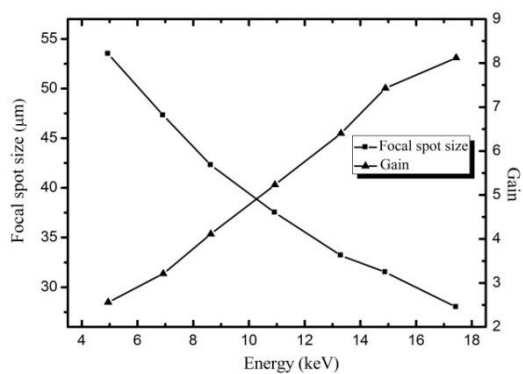

Fig. 2. The IFS size and the gain of the PPXRL.

The OFS size and the gain of the PFXRL were 30.1 μm and 3400 at 17.4 keV, respectively. Fig. 2 shows the IFS size and the gain of the PPXRL. The gain of the PPXRL was measured using a pinhole with a diameter of 500.0 μm [25]. The IFS size of the PPXRL was measured using an X-ray source scan [25]. The decrease of the IFS size of the PPXRL resulted from the decrease of the critical angle of total reflection with higher energies.

The profile size of the confocal micro-volume depended on the OFS size, $d_o$, of PFXRL and the IFS size, $d_i$, of PPXRL. When $\phi = \psi = 45°$ (Fig. 1), the size of the profile of the confocal micro-volume could be written as following formulas:

$$l_x = l_y = \frac{d_i + d_o}{\sqrt{2}}. \tag{1}$$

$$l_z = min\{d_i, d_o\}. \tag{2}$$

where $l_x$, $l_y$ and $l_z$ are the profile size of the confocal micro-volume along the x, y and z axis (Fig. 1), respectively [26]. The energy dependence of $d_i$ was shown in Fig. 2, and $d_o$ was 30.1 μm at 17.4 keV. Because of the use of quasi-monochromatic X-rays in exciting channel, the energy of the exciting X-rays was about 17.4 keV which was higher than that of the XRF of the element from the sample. This resulted in that the sizes of different $d_i$ corresponding to various XRF of elements from sample were larger than that of $d_o$ which was 30.1 μm. Therefore, $l_z$ =30.1 μm and remained unchanged while both $l_x$ and $l_y$ reduced with the increasing energies.

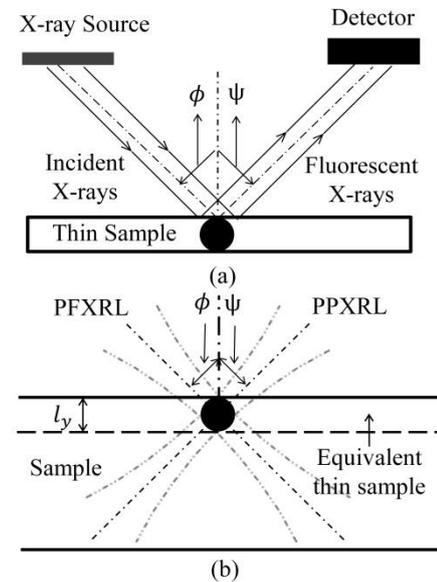

Fig. 3. (a) The sketch of conventional XRF with thin sample technology. (b) The sketch of confocal XRF with the CETSM.

**2.2 Equivalent thin sample method for quantitative analysis based on confocal micro-volume**

Fig. 3a shows the typical arrangement of XRF. The specimen can be seen as thin sample with a mass per unit area, m, as following [27]:

$$m < \frac{0.1}{\mu(E_o)\mathrm{cosec}\phi + \mu(E_i)\mathrm{cosec}\psi} . \quad (3)$$

where $\mu(E_o)$ and $\mu(E_i)$ are the total mass attenuation coefficients for the whole specimen at the energy of primary radiation ($E_o$) and the energy of characteristic X-rays of the ith element ($E_i$), respectively, $\phi$ is the effective angle of incidence of the primary exciting beam, and $\psi$ is the effective take-off angle of characteristic X-rays. The total mass attenuation coefficient $\mu(E)$ for the whole specimen at the energy $E$ is given by the mixture rule:

$$\mu(E) = \sum_{j=1}^{n} W_j \mu_j(E). \quad (4)$$

where $W_j$ and $\mu_j(E)$ are the weight fraction and the mass attenuation coefficient of the $j^{th}$ element present in the sample, respectively, and n is the total number of the elements in the sample.

In conventional XRF quantitative analysis, if a homogeneous sample to be analysis satisfies the Eq. (3), the intensity of characteristic X-rays, $I_i$, of the i element can be simply calculated by [27]:

$$I_i = S_i m_i. \quad (5)$$

$$S_i = \frac{G}{\sin\phi} I_i(E_o)\varepsilon(E_i)\tau_i(E_o)\omega_i p_i \left(1 - \frac{1}{j_i}\right) . \quad (6)$$

$$m_i = \mu_i m. \quad (7)$$

where $G$ is the geometry factor, $\phi$ is the effective incidence angle for primary radiation, $I_o(E_o)$ is the intensity of primary photons of energy $E_o$ (monochromatic excitation), $\varepsilon(E_i)$ is the detector efficiency for recording the photons of energy $E_i$, $\tau_i(E_o)$ is photoelectric mass absorption coefficient for the ith element at the energy 0, in $cm^2 \cdot g^{-1}$, $\omega_i$ is the fluorescence yield of the element i, $p_i$ is the transition probability of the k[th] line of the element i, and $j_i$ is the absorption jump at the K-edge of photoelectric absorption in ith element.

In this paper, a quantitative method for analyzing the intermediately and infinitely thick samples using thin sample approach based on the confocal micro-volume without sample preparation was developed. Compared with the conventional XRF analysis (Fig. 3a), when the confocal micro-volume was just entered into the surface of the sample (Fig. 3b), only the sample in the confocal micro-volume was analyzed. At this condition, the confocal XRF could be seen as an analysis of a thin layer sample with a thickness roughly equal to the profile size, $l_y$, of the confocal micro-volume. The thickness of the thin layer sample restricted in the confocal micro-volume could meet the requirement of the thin sample method. We called this thin sample method based on the confocal micro-volume as the "confocal equivalent thin sample method (CETSM)".

As mentioned above, as for the thin sample method, the notable feature is that the absorption and enhancement effects can be neglected. In other word, the concentration of ith element in sample is proportional to its intensity of characteristic X-rays. The sensitivity factor $S_i$ from Eq. (5), which is a constant for a specific element in the sample, is used to convert the intensity of characteristic X-rays into its concentration. The sensitivity factor $S_i$ for a specific element $i$ can be measured experimentally as the slope of the straight line obtained by determining a group standard sample with a different concentration of element $i$. Using this method, the sensitivity factor curve for different elements could be obtained.

## 3 Results and discussions

In order to verify the effectiveness of the CETSM, ions in solution were analyzed. According to the Eq. (3) and the size of confocal micro-volume, when the whole confocal micro-volume was immersed in the specimen in our experiment, the analyzed layer sample restricted in the confocal micro-volume could be disposed as equivalent thin sample for the element from Co to Y, and, in other words, such elements could be analyzed with the CETSM. When the confocal micro-volume was partly put into the liquid sample by 15 μm, the CETSM could be used to analyze the elements of Cr, Mn and Fe in our study.

Putting the confocal micro-volume in the appropriate place within the sample is important for reducing the errors of quantitative analysis with the CETSM. The confocal micro-volume might be put at the right position within the sample as following. When the surface of the analyzed sample scanned through the confocal micro-volume, the integral and differential distribution of the XRF intensity of $Cu - K_\alpha$ from the sample was shown in Fig. 4. The Gaussian-distribution of XRF in the differential curve (Fig. 4) resulted from the Gaussian-distribution of the exciting X-rays from the PFXRL in the confocal micro-volume. The full width at half maximum (FWHM) of the differential curve was about 52.5 μm, which agreed with the value calculated by Eq. (1) base on $d_i$ =44.5 μm for $Cu - K_\alpha$ at 8.0 keV in Fig. 2 and $d_o$ = 30.1 μm at 17.4 keV. The center, C, of the differential curve (Fig. 4) was corresponding to the part of the scan when the surface of the sample located at the center of the confocal micro-volume of the setup. The profile size of the confocal micro-volume was known. Therefore, the confocal micro-volume could be put the right position within the sample by such scans.

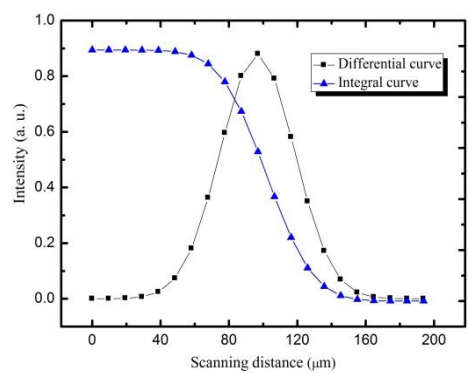

Fig. 4. Integral and differential distribution of the XRF intensity of the studied element from the sample.

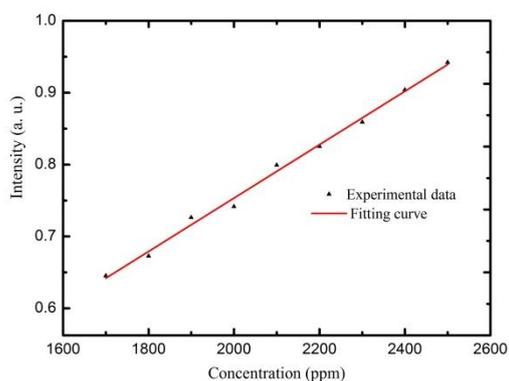

Fig. 5. The standard curve of the Cr-Kα fluorescence.

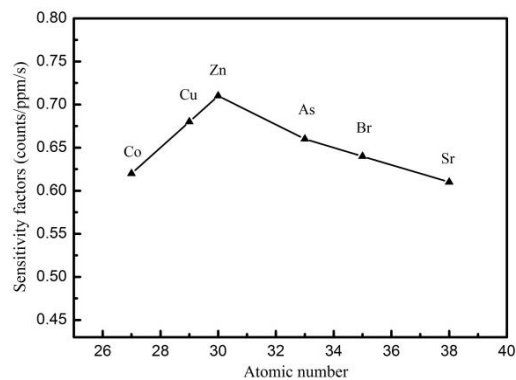

Fig. 6. Sensitivity factors for corresponding elements.

To acquire the sensitivity factors for each studied element, a series of standard solutions were prepared, respectively. With Cr for example, the standard curve of the $Cr-K_\alpha$ fluorescence, which is the intensity corresponding to a series concentration of $CrCl_3$, is shown in Fig. 5, and then the sensitivity factor of Cr could be acquired as the slope of the fitting straight line. With the same method, the sensitivity factors of other elements for the CETSM also could be obtained. Fig. 6 showed the sensitivity factor for the elements of Co，Cu, Zn, As, Br and Sr, which were measured with the whole confocal micro-volume just immerged inside the surface of the sample. As for Cr, Mn and Fe, their sensitivity factors were 0.11, 0.15 and 0.19counts/ppm/s, respectively, which were measured with the confocal micro-volume partly immerged inside the surface of the sample by 15 μm. For the sensitivity factors shown in Fig. 6, although they were obtained with the whole confocal micro-volume just immerged inside the surface of the sample, they could be used to analyze the sample with the confocal micro-volume located at any positions inside the sample with the CETSM. For example, the detected XRF intensity for the sample in the confocal micro volume which was placed inside the sample by depth of t (Fig. 7) could be corrected into the equivalent intensity with the confocal micro volume just immerged inside the surface of the sample by the following equation:

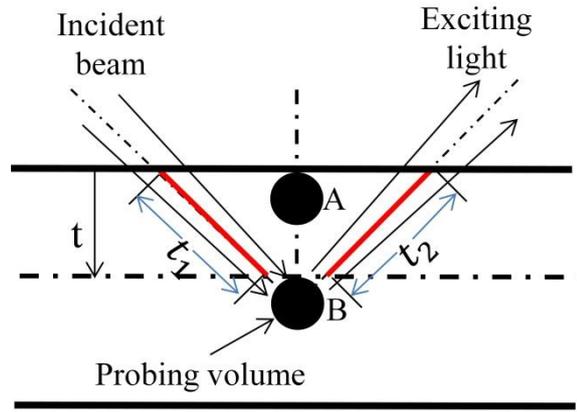

Fig. 7. The sketch of absorption correction of the sensitivity factor in a certain depth of the sample.

$$I_{real} = \frac{I_{detected}}{e^{-(\mu_1 t_1 + \mu_2 t_2)}} \tag{8}$$

where $\mu_1$ was the absorption coefficient from the sample for the exciting X-rays from the PFXRF, $\mu_2$ was the absorption coefficient from the sample for the excited XRF from the confocal micro-volume. Therefore, when the analyzed elements met the requirement of the CETSM when the whole confocal micro-volume was immerged inside samples, the three-dimensional information about such elements could be obtained by using the CETSM.

To demonstrate the effectiveness of the CETSM, a solution prepared with multiple ions was analyzed. The working voltage and current of the X-ray source were 30 kV and 40 mA, respectively. The detection time was 70 seconds for every point. The element concentrations determined with the CETSM were shown in Table 1 with the whole confocal micro volume just immerged inside the surface of sample for the Cu, As and Sr, and with the confocal micro volume partly placed inside the sample by 15 μm for Cr and Fe. The relative errors of this method in determining the concentration of the ions in solution were better than 8%. Table 2 shows the concentrations of the element Cu, As and Sr in the same solution detected by putting the confocal micro-volume in the depth 50 μm, namely $t$ =50 μm in Fig. 7, of the sample and the relative errors were better than 10%.

Table 1. Concentrations of the elements detected by the CETSM.

| Elements | Cr | Fe | Cu | As | Sr |
| --- | --- | --- | --- | --- | --- |
| True value (ppm) | 2540 | 1850 | 1550 | 1350 | 2850 |
| Experimental value (ppm) | 2632 | 1747 | 1587 | 1299 | 2934 |
| Relative error (%) | 7.43% | 5.57% | 2.39% | 3.78% | 2.95% |

Table 2. Concentrations of the elements detected by putting the confocal micro-volume inside the surface of the sample by 50 μm.

| Elements | Cu | As | Sr |
| --- | --- | --- | --- |
| True value (ppm) | 1550 | 1350 | 2850 |
| Experimental value (ppm) | 1398 | 1456 | 3025 |
| Relative error (%) | 9.81% | 7.85% | 6.14% |

In the experiment, the liquid sample was put in the plastic bag. The nonplanar surface of the bag might produce errors of the experimental results.

As shown in Fig. 8, different elements had different corresponding confocal micro-volumes. As mentioned above, a series of standard solution for one element were prepared and used to acquire the sensitivity factor of this element. However, for the elements with a larger atomic number than that of Co, their sensitive factors were measured with the whole confocal micro-volume for the element of Co just immersed inside the surface in the standard solution, respectively. The reason for this was as following. For such sensitive factors, when they were used to determine simultaneously the densities of multiple elements at any position inside the sample, the determination of $t_1$ and $t_2$ in the Fig. 7 were same for different elements. Therefore, such sensitive factors were convenient for determining multi-elements simultaneously.

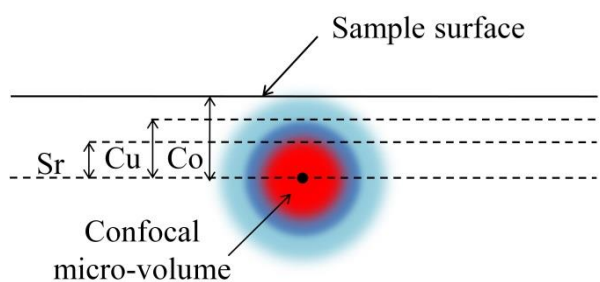

Fig. 8. Confocal micro-volume for different elements.

According to Eq. (3), the size of confocal volume had a significant influence on the application fields of the CETSM in this paper. As mentioned above, with the PFXRL and the PPXRL used in our experiments, in order to analyze the elements of Cr, Mn and Fe with the CETSM, the confocal micro-volume was partly put into the liquid sample by 15 μm which was smaller than the profile size of the confocal micro-volume. Therefore, with the PFXRL and the PPXRL used here, the CETSM could only be used in analyzing the surface of thick sample or a thin film sample with such elements as Cr, Mn and Fe. Moreover, that only part of confocal volume could be used by the CETSM resulted in the waste of the most of the X-rays from the PFXRL. In our experiments, although the elements with a larger atomic number than that of Co could be analyzed by the CETSM with the whole confocal micro-volume used, the CETSM could only be used to determine the low concentration of heavy elements in light matrices with the PFXRL and the PPXRL used in our experiments. When the profile size of the confocal micro-volume was smaller, it was easier to meet the requirement of the thin sample method with the CETSM when the confocal micro-volume was fully immersed in the sample. Therefore, there were more application fields with a smaller confocal micro-volume. Because the profile size of the confocal micro-volume depended on the size of the foci of the X-ray optics, a confocal XRF setup with a smaller confocal micro-volume could be obtained with X-ray optics with smaller foci, such as Kirkpatrick-Baez mirrors, compound refractive lenses, monocapillary, and so on [10, 11, 28, 29].

**4 Conclusions**

The confocal XRF based on a PFXRL and a PPXRL was used to quantitatively analyze multi-element solutions in order to validate that the confocal XRF had potential applications in analyzing the intermediately and infinitely thick samples with thin sample approach without sample preparations. The relative errors of the CETSM in determining the concentration of the ions in solution were better than 10%. The CETSM need neither the multifarious specimen pretreatment compared with conventional XRF thin sample technology nor the complicated calculations compared with the other quantitative analysis method with confocal XRF setup. The CETSM had potential applications in analyzing the intermediately and infinitely thick samples with thin sample approach without sample preparations.